\documentclass[10pt,a4paper,notitlepage]{revtex4-1}
\usepackage{graphicx}
\usepackage{amsmath}
\usepackage{subcaption}
\usepackage{color}
\usepackage{fullpage}

\begin{document}
\title{Controlled processing of signal stored in metamaterial with tripod structure}
\author{S. Zieli\'nska-Raczy\'nska}
\author{D. Ziemkiewicz}
\email{david.ziemkiewicz@utp.edu.pl}
\affiliation{Institute of Mathematics and Physics, UTP University of Science and Technology, Al. Kaliskiego 7, 85-789 Bydgoszcz, Poland.}

\begin{abstract}
 In the present paper we have discussed in detail electromagnetically induced
transparency and signal storing in the case of one signal
pulses propagating in a classical electric medium resembling this of four-level atoms in the
tripod configuration. Our theoretical results confirm  recently observed dependence of transparency windows position on coupling parameters. In the process of storing the pulse energy is
 confined inside the metamaterial as electric charge oscillations  and after required  time  it is possible to switch the control fields on
again and to release the trapped signal.  By manipulating the driving fields one can thus control
the parameters of the  released signal and even to divide it on demand into arbitrary parts.

\end{abstract}
\maketitle %% required
\section{Introduction}

 A striking and important example of the phenomenon which completely alters the conditions for propagation of the electromagnetical waves through a medium is the  electromagnetically induced transparency (EIT) \cite{Harris}-\cite{RMP} which consist in making the medium transparent for a pulse resonant with some atomic transition by switching on a strong  field resonant with two unpopulated levels. Generic EIT scheme consists of a gas comprised of three-level atoms in so-called $\Lambda$ configuration that are driven by a strong control field and a weak probe one on separate transition. If the frequencies of probe and control fields are in two-photon resonance the atoms are driven into a dark-state that is decoupled from the light fields. This mechanism creates the transparency window in the absorption spectrum of the probe field irradiating the medium, and the steep normal dispersion at the center of the dip of the absorption and  results in significant reduction of the group velocity of the signal and enhancement of nonlinear interaction.  By admitting the control fields to adiabatically change in time it has become possible to dynamically change the optical properties of the medium while the probe pulse travels inside it \cite{Andre}; in particular one can not only reduce the pulse group velocity but even stop the pulse by switching the control field off. By switching it again one can release the stored pulse, preserving the phase relations.
 During the last twenty years or so, EIT and other related to its phenomena were studied both theoretically and experimentally and more and more sophisticated variation of atomic configurations and coupling fields have been considered \cite{2lambda}, \cite{Wu}.

A tripod-type four-level configuration, the classical analog of which will be considered in this paper, consists of three ground levels  coupled with one upper level. Even at a first glance one can see that this configuration, due to an additional unpopulated long-living lower state coupled by control or second signal beam with upper state, is richer than generic $\Lambda$ scheme and gives an opportunity to study different new aspect of pulse propagation. We will focus on the situation of one signal coupling populated ground level with the upper level and two strong control fields coupled with two low-lying empty levels.  This medium  exhibits in general two transparency windows of different widths and different slopes of normal dispersion curve \cite{Paspalakis}, what means that the group velocity in the two windows can be different. Extensive studies of possibilities of stopping, performing manipulations on stored classical light in order to obtain the desired properties of the released pulse or pulses were done by one Raczy\'nski \emph{et al} \cite{Malgosia}. Medium of atoms in tripod configuration, due to high degree of controllability, may also play the role of a light-controlled beam-splitting system in the time domain, allows one to realize the Hong-Ou-Mandel interference, one-and two-qubit gates, working on atomic excitations due to stores polarized photons and processes time entangled pulses (light bins) \cite{HOM}, \cite{Karolina}, \cite{statystyka}.   Recent experiments by Wang \emph{et al} \cite{Wang} and Yang \emph{et al} \cite{Yang} realized controllable splitting and modulation of classical pulse and single-photon-level pulses in a tripod-type atomic medium.

EIT has turned out to be useful tool in quantum optics providing realization of various kinds of sophisticated experiments such as mentioned above light stopping and storing, cross-Kerr or cross-phase modulation \cite{RMP}.  Nowadays a lot of attention has been paid to classical analog of EIT  metamaterials \cite{Souza}, \cite{my2016}, where one expects to perform similar experiments taking advantage of operating at room temperatures.
Moreover, in metamaterials consisting of coupled split-ring resonators, no quantum mechanical atomic states are required to observe EIT which may lead to slow light applications in a wide range of frequency, from microwave up to the terahertz \cite{Tassin2009} and infared \cite{liu} regime. Recently the storage of a signal in metamaterial has been demonstrated  \cite{Nakanishi}.
The investigations of classical analogs of EIT media has been motivated by recognition of wide bandwidth, low loss propagation of signal through initially thick media, opening many prospects on novel optical components such as tunable delay elements, highly sensitive sensors and nonlinear devices. Our paper is a contribution to this area.
 Some aspects of the atom-field interaction can be described by classical theory  so we attempt to present the classical systems that mimic tripod configuration and explore the signal storage in such  media.  As it was pointed out, tripod scheme allows one to perform manipulations on the signal pulse stopped in the medium, and after some time to release it in one or more parts, preserving information of the incident field, including its amplitude, phase and  polarization stage.  A classical scheme that mimics tripod has been investigated recently by Bai \emph{et al} \cite{Bai2013} in the context of double EIT and plasmon-induced transparency  \cite{Bai2016}, but all these papers considered two weak propagating fields coupling populated low-lying levels in the medium dressed by one control beam. Studies of EIT in metamaterials of plasmonic tripod system has also been performed by Xu \emph{et al} \cite{Xu2}, who considered the off resonant situation of all three fields.
 We investigate the EIT in classical tripod medium built of three RLC circuits coupled  by two capacitances with electric resistors and alternating voltage source, or alternatively by several metal strips. Our model could be practically realized using planar complementary metamaterials in which generic EIT was sucesfully demonstrated  by Li \emph{et al}  \cite{liu}.
 The electric susceptibility of such a system allows one for opening two transparency windows for an incident external field, with their width depending on system geometry and coupling capacitances.

 Our paper is organized as follows. In the first section, the classical model of atomic tripod system is discussed and the means of control over the medium dispersion are explained. Then, the application of Finite Difference Time Domain (FDTD) method to the simulation of pulse propagation through EIT metamaterial is described. Finally, the simulation results are presented, including verification of the medium model and simulation of signal stopping and its release as a single part or in a series of pulses.

\section{Classical analogues of an atomic tripod system}
\subsection{The model}

\begin{figure}[ht!]
    \centering
    \begin{subfigure}[b]{0.4\linewidth}
    \includegraphics[width=1\linewidth]{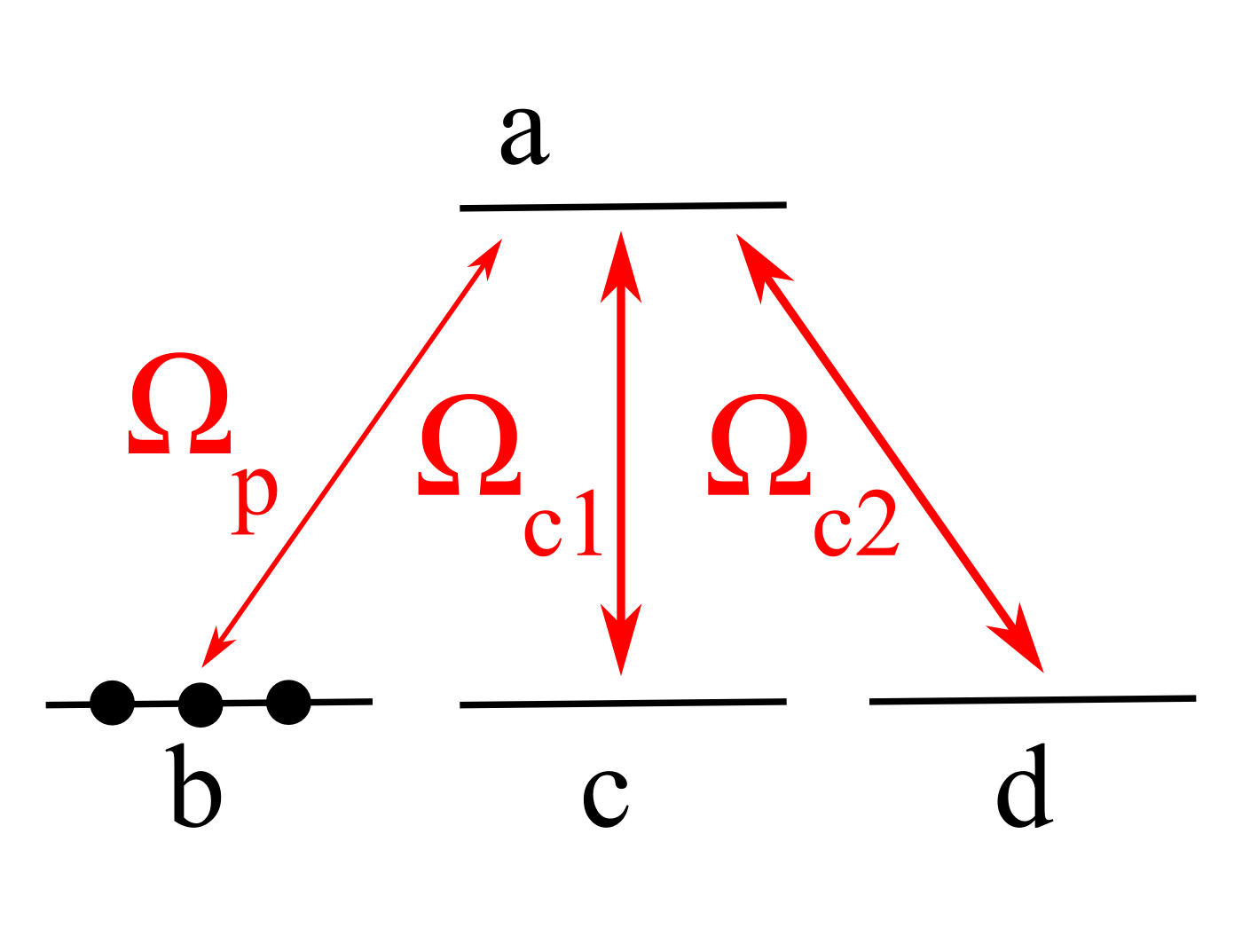} \caption{} \label{rysEIT}
    \end{subfigure}
    \begin{subfigure}[b]{0.4\linewidth}
    \includegraphics[width=1\linewidth]{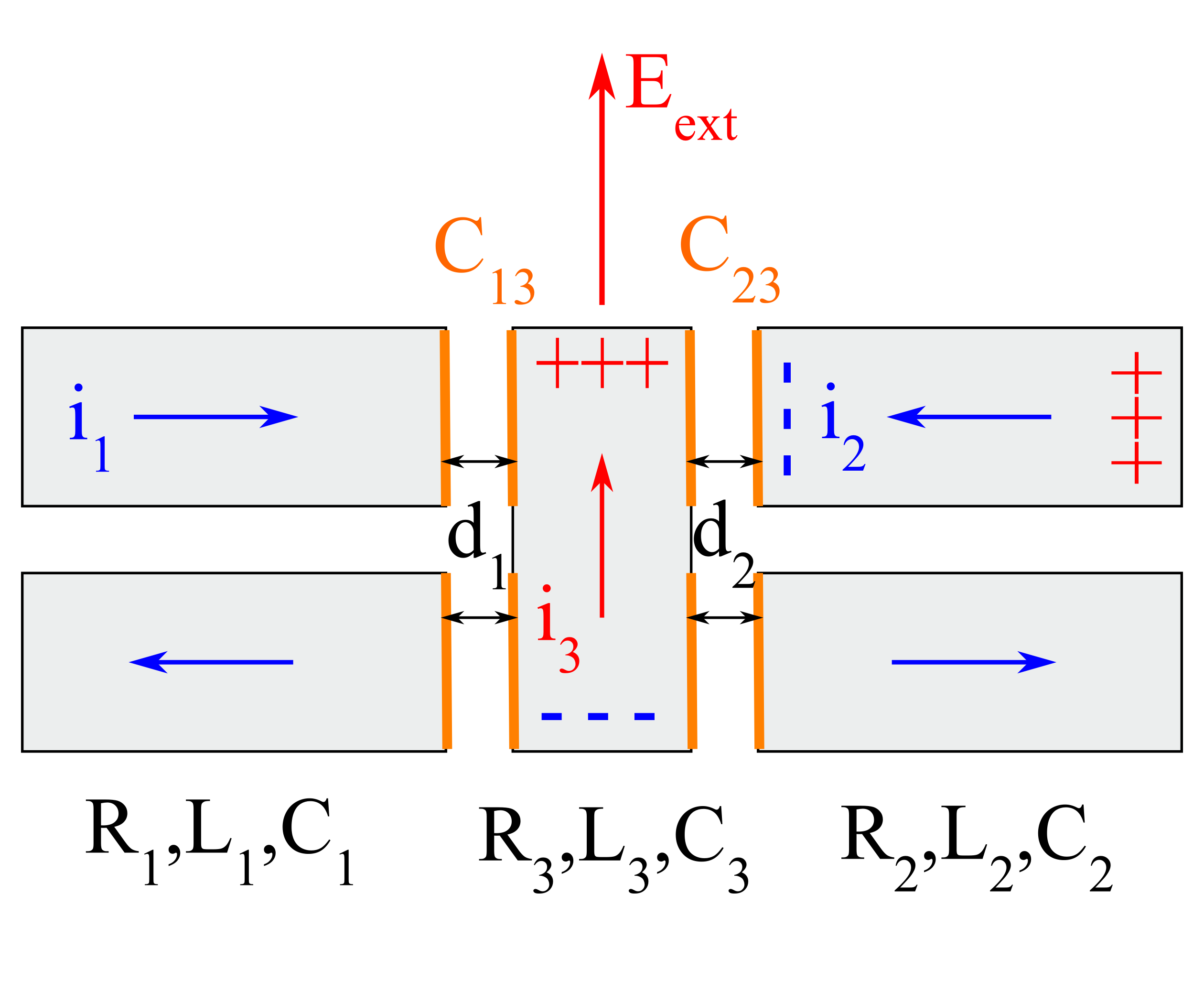} \caption{} \label{uklad}
    \end{subfigure}
    \begin{subfigure}[b]{0.4\linewidth}
    \includegraphics[width=1\linewidth]{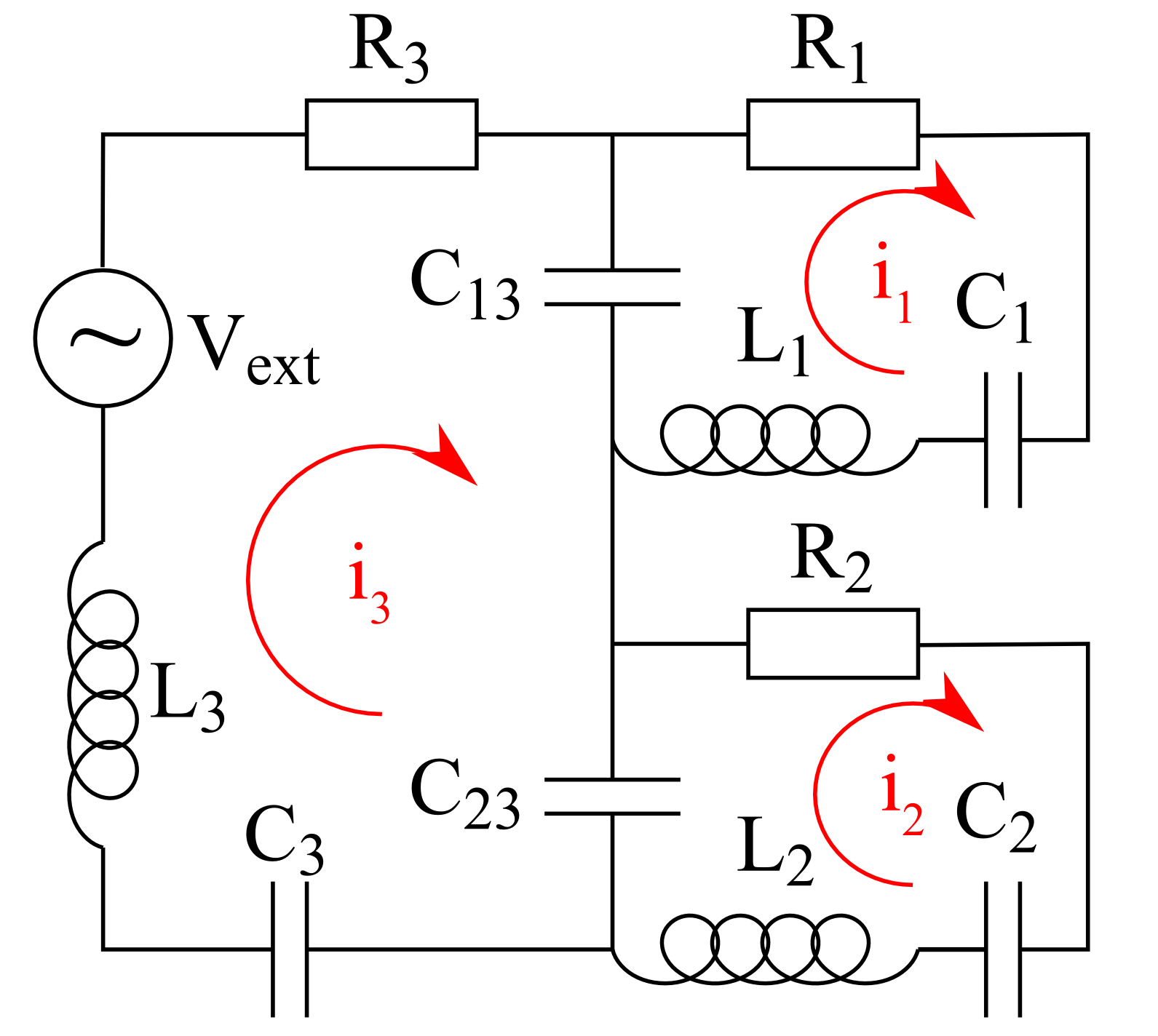} \caption{} \label{circuit}
    \end{subfigure}
    \begin{subfigure}[b]{0.4\linewidth}
    \includegraphics[width=1\linewidth]{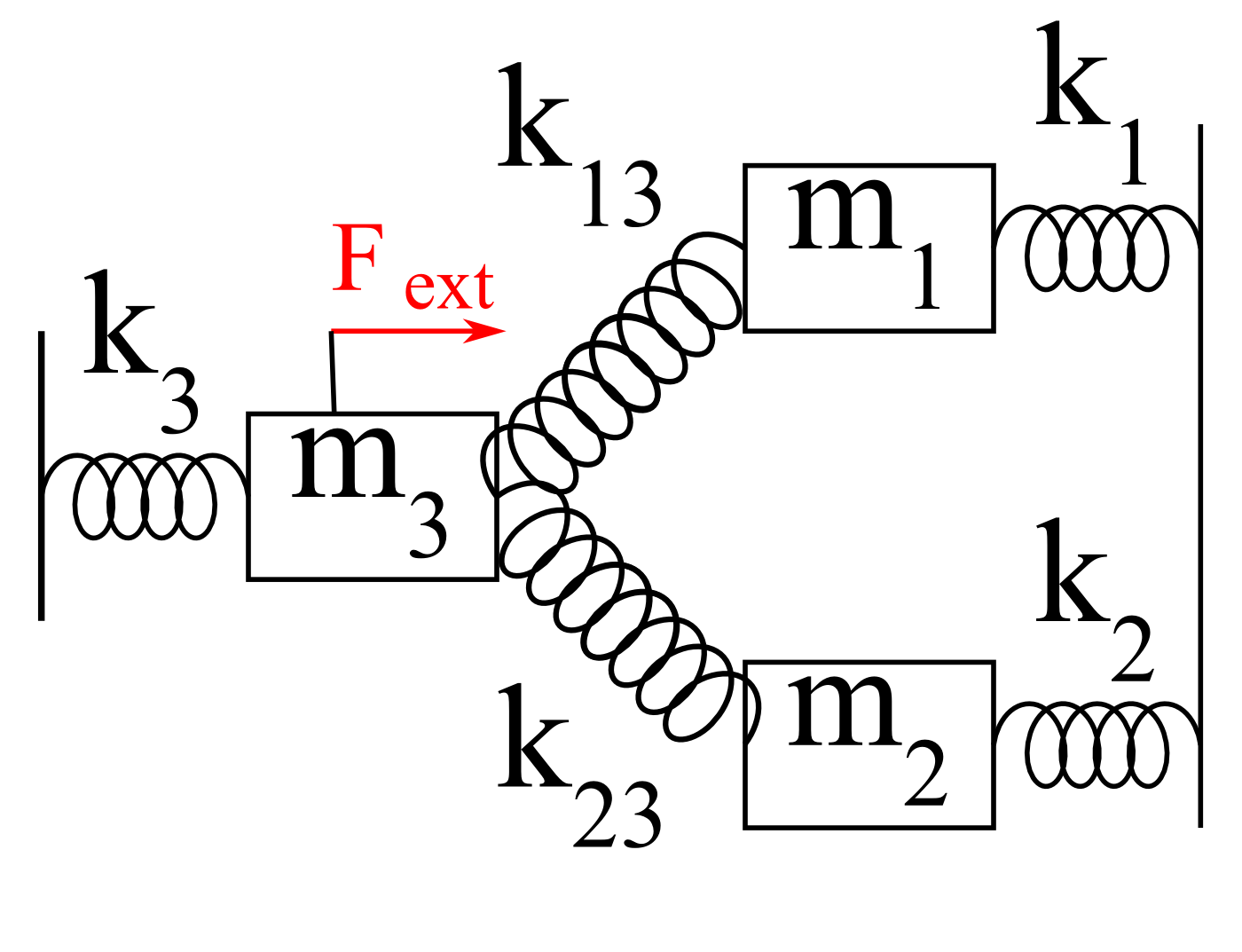} \caption{} \label{mech}
    \end{subfigure}
    \caption{a) Schematics of a three level tripod configuration with a single probe field $\Omega_p$ and two control fields $\Omega_c1$, $\Omega_c2$ b) Metal strip antenna with single radiative element and two dark resonators c) Electric circuit model of the system d) The mechanical analogue of tripod system}
\end{figure}
The electric analog of the atomic system in tripod configuration with a single probe and two strong control fields presented in Fig. 1a is
 shown in Fig. \ref{uklad}.  There is one radiative element with current $i_3$ caused by the external electric field $E_{ext}$. Moreover, there are two additional elements not interacting with the external field directly \cite{Xu2}.
  The charges accumulating at the ends of the central slab induce currents $i_1$ and $i_2$. This coupling, represented by capacitances $C_{13}$ and $C_{23}$, depends on the spatial distance between the slabs, marked by $d_1$ and $d_2$.
 In the case of a long, thin wire, the inductance $L$ introduces additional inertia to the system, changing the effective mass of the electrons \cite{Pendry}. Moreover, when the wire has a finite length, the motion is constrained and the accumulation of charge on the ends acts as a capacitor and provides a restoring force, so that the slab acts as an electric, harmonic oscillator.
Therefore, as it was pointed out in \cite{Nowotny}, a metal strip operates as an optical dipole antenna, with intrinsic resistance $R$, inductance $L$ and capacitance $C$ dependent on its geometry. By placing two metal strips  parallel to each other, one can form a so-called dark resonator \cite{Zhang2008} with no dipole coupling to the external field due to the counterpropagating currents.  The electric circuit model of the system is shown on the Fig. \ref{circuit}. The currents in the circuit are described by the system of equations
\begin{eqnarray}
V_{ext}&=&L_3 \frac{\partial i_3}{\partial t} + R_3 i_3 + \frac{1}{C_3}\int i_3 dt + \frac{1}{C_{13}}\int (i_3-i_1) dt + \frac{1}{C_{23}}\int (i_3-i_2) dt, \nonumber\\
0 &=& L_1 \frac{\partial i_1}{\partial t} + R_1 i_1 + \frac{1}{C_1}\int i_1 dt - \frac{1}{C_{13}}\int (i_3-i_1) dt, \nonumber\\
0 &=& L_2 \frac{\partial i_2}{\partial t} + R_2 i_2 + \frac{1}{C_2}\int i_2 dt - \frac{1}{C_{12}}\int (i_3-i_2) dt.
\end{eqnarray}
In terms of an electric charge $q$, where $\dot{q}=i$ the above set of equations takes the form
\begin{eqnarray}
V_{ext}&=&L \ddot{q_3} + R_3 \dot{q_3} + (\frac{1}{C_3}+\frac{1}{C_{13}}+\frac{1}{C_{23}})q_3 - \frac{1}{C_{13}}q_1 - \frac{1}{C_{23}}q_2, \nonumber\\
0 &=& L \ddot{q_1} + R_1 \dot{q_1} + (\frac{1}{C_1}+\frac{1}{C_{13}})q_1 - \frac{1}{C_{13}}q_3, \nonumber\\
0 &=& L \ddot{q_2} + R_2 \dot{q_2} + (\frac{1}{C_2}+\frac{1}{C_{23}})q_2 - \frac{1}{C_{23}}q_3.
\end{eqnarray}
where, without loss of generality, $L_1=L_2=L_3=L$ was assumed. Finally, we arrive at the equations for three coupled, harmonic electric oscillators
\begin{eqnarray}\label{diff_eq}
\ddot{q_3} &+& \gamma_3 \dot{q_3} + \omega_3^2 q_3 - \Omega_{1}^2 q_1 - \Omega_{2}^2 q_2 = \frac{V_{ext}}{L}, \nonumber\\
\ddot{q_1} &+& \gamma_1 \dot{q_1} + \omega_1^2 q_1 - \Omega_{1}^2 q_3  = 0, \nonumber\\
\ddot{q_2} &+& \gamma_2 \dot{q_2} + \omega_2^2 q_2 - \Omega_{2}^2 q_3  = 0,
\end{eqnarray}
where
\begin{eqnarray}\label{param_elec}
\omega_1^2 &=& (\frac{1}{LC_1}+\frac{1}{LC_{13}}) \qquad \gamma_1 = \frac{R_1}{L}, \nonumber\\
\omega_2^2 &=& (\frac{1}{LC_2}+\frac{1}{LC_{23}}) \qquad \gamma_2 = \frac{R_2}{L}, \nonumber\\
\omega_3^2 &=& (\frac{1}{LC_3}+\frac{1}{LC_{13}}+\frac{1}{LC_{23}}) \qquad \gamma_3 = \frac{R_3}{L}, \nonumber\\
\Omega_{1}^2 &=& \frac{1}{LC_{13}} \qquad \Omega_{2}^2 = \frac{1}{LC_{23}}.
\end{eqnarray}
The presented circuit and obtained system of equations is analogous to the one described in \cite{Bai2013}, with the exception that in our case, only one lower level is populated, so a single probe field $E_{ext}$ generating potential $V_{ext}$ is assumed. Analogous set of equations can be obtained for displacements $x_i$ in a system of coupled, mechanical oscillators (Fig. \ref{mech}) and it is a straightforward extension of the models for a lambda EIT system \cite{Alzar,Tassin}.

 The potential  $V_{ext}=V_0 \exp(-i\omega t)$ generated by external field $E_{ext}=E_0\exp(-i\omega t)$ causes the charge oscillations in the form $q_i=q_{0i}\exp(-i\omega t)$ for $i=1,2,3$. Substituting these expressions into Eq. \ref{diff_eq}, we are able to solve for the steady state solution in the form
\begin{equation}\label{dyspersyjna}
q_{03}(\omega) = \frac{1/L}{\omega_3^2 - \omega^2 - i\gamma_3\omega - \frac{\Omega_1^4}{\omega_1^2 - \omega^2 - i\gamma_1\omega} - \frac{\Omega_2^4}{\omega_2^2 - \omega^2 - i\gamma_2\omega}}V_0
\end{equation}
The equation links the charge $q_{03}$ with the potential $V_0$ and is expressed in units of capacitance; $q_{03}(\omega) = C_{eff}(\omega)V_0$ where $C_{eff}$ is an effective capacitance of the circuit. Suppose that the circuit is a model of a metallic structure of the length $d$ reacting to the external electric field $E_{ext}$. Then, the potential is $V_0 = E_0d$ and the induced charge generates a dipole moment $P=q_{03}d$. Thus we can write
\begin{eqnarray}
q_{03}d = C_{eff}(\omega)E_0d^2 \nonumber\\
P(\omega) = d^2 C_{eff}(\omega)E_0 \nonumber
\end{eqnarray}
so the susceptibility, which in general is a complex function, is given by
\begin{equation}\label{susc}
\chi(\omega) = \chi'+i\chi''=\frac{d^2/\epsilon_0 L}{\omega_3^2 - \omega^2 - i\gamma_3\omega - \frac{\Omega_1^4}{\omega_1^2 - \omega^2 - i\gamma_1\omega} - \frac{\Omega_2^4}{\omega_2^2 - \omega^2 - i\gamma_2\omega}}.
\end{equation}
\begin{figure}
\centering
\includegraphics[width=0.5\linewidth]{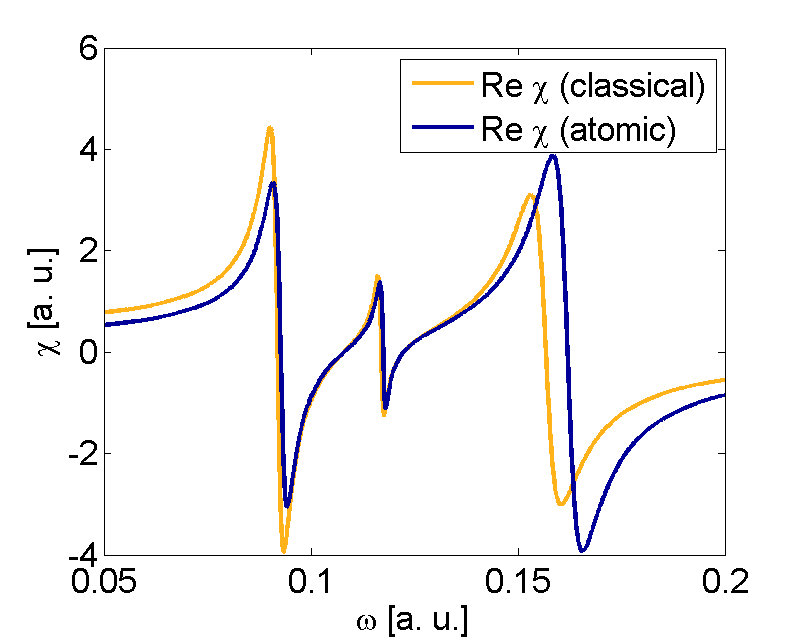}
\caption{The real part of $\chi(\omega)$ for the medium parameters $L=100$, $R_1=R_2=0.1$, $R_3=1$, $C_1=C_3=1$, $C_2=1.5$, $C_{13}=C_{23}=2$.} \label{rys_dysp}
\end{figure}
In the case of low detuning, very close to the resonance, when $\omega_1 \approx \omega_2 \approx \omega_3 \approx \omega$, we have \hbox{$\omega_i^2 - \omega^2 \approx 2\omega(\omega_i - \omega)$} for $i=1,2,3$. Introducing new quantities
\begin{eqnarray}\label{atom_param}
\Omega_{c1}^2 = \frac{\Omega_1^4}{4 \omega^2} \approx \frac{\Omega_1^4}{4 \omega_1^2} = \frac{C_1}{4LC_{13}(C_1+C_{13})} \nonumber\\
\Omega_{c2}^2 = \frac{\Omega_2^4}{4 \omega^2} \approx \frac{\Omega_2^4}{4 \omega_2^2} = \frac{C_2}{4LC_{23}(C_2+C_{23})} \nonumber\\
\gamma_{ab} = \frac{\gamma_3}{2}, \qquad \gamma_{bc} = \frac{\gamma_1}{2}, \qquad \gamma_{bd} = \frac{\gamma_2}{2},
\end{eqnarray}
it is possible to obtain the dispersion relation for the atomic tripod system
\begin{equation}\label{susc_A}
\chi(\omega) = \frac{-A}{\omega - \omega_3 + i \gamma_{ab} - \frac{\Omega_{c1}^2}{\omega - \omega_1 + i\gamma_{bc}} - \frac{\Omega_{c2}^2}{\omega - \omega_2 + i\gamma_{bd}}}
\end{equation}
where $A$ is a positive constant. Note, that setting $\Omega_{c2}=0$ in Eq. \ref{susc_A}, one gets the  dispersion relation for a three level lambda system.
For a while we would like to concentrate on  susceptibilities  given by Eq. \ref{susc} and \ref{susc_A}, describing the classical electric analog of EIT system and the quantum system, respectively. It can be seen from the Fig. \ref{rys_dysp} there is a close correspondence between both susceptibility values   for the most important frequency region inside the transparency window and normal dispersion. Small discrepancy is  noticeable only outside the transparency windows, in the regions of anomalous dispersion and high absorption.

\subsection{Steering of dispersion in classical tripod medium}
\begin{figure}[ht!]
\centering
    \begin{subfigure}[b]{0.48\linewidth}
    \includegraphics[width=1\linewidth]{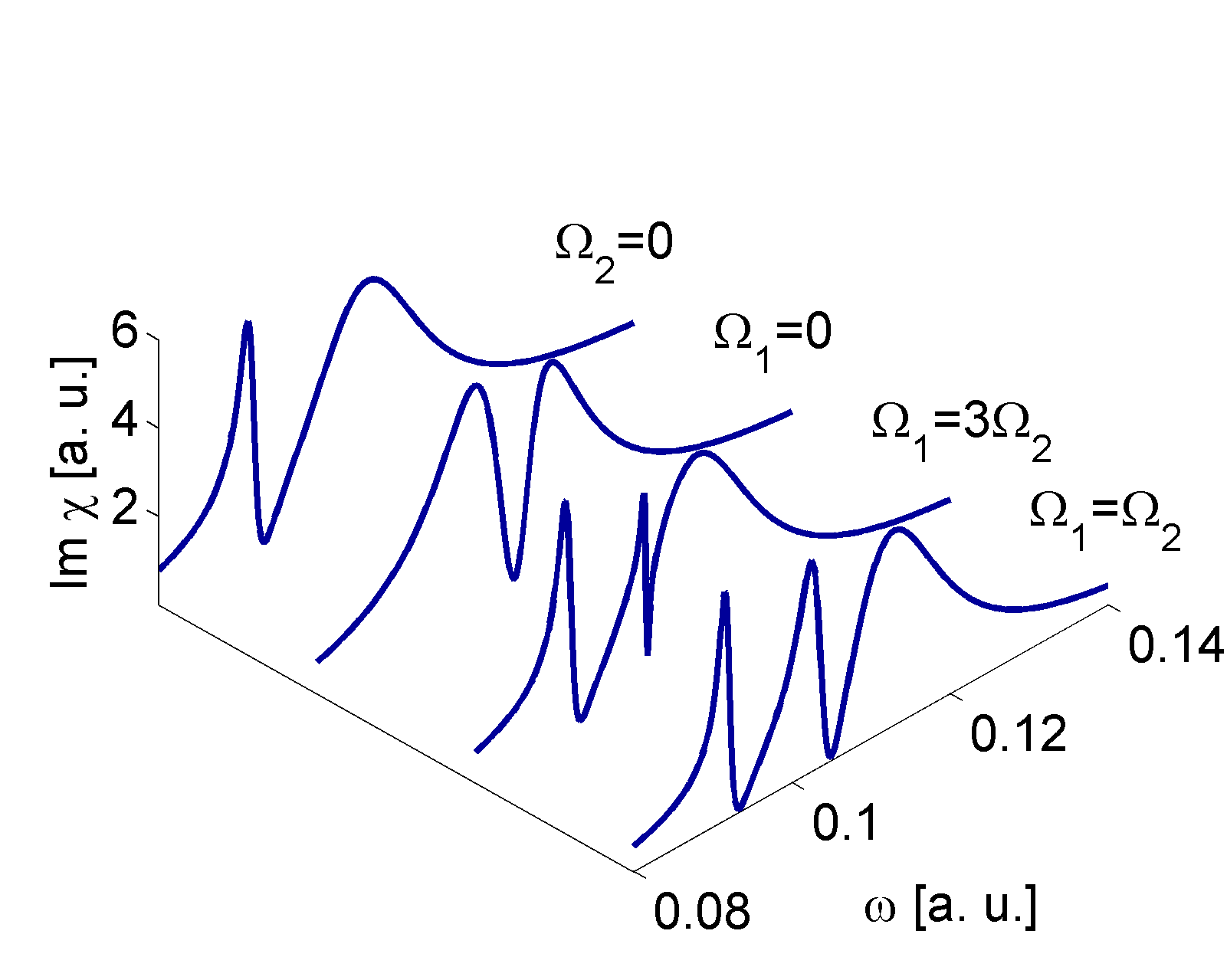} \caption{}
    \end{subfigure}
    \begin{subfigure}[b]{0.48\linewidth}
    \includegraphics[width=1\linewidth]{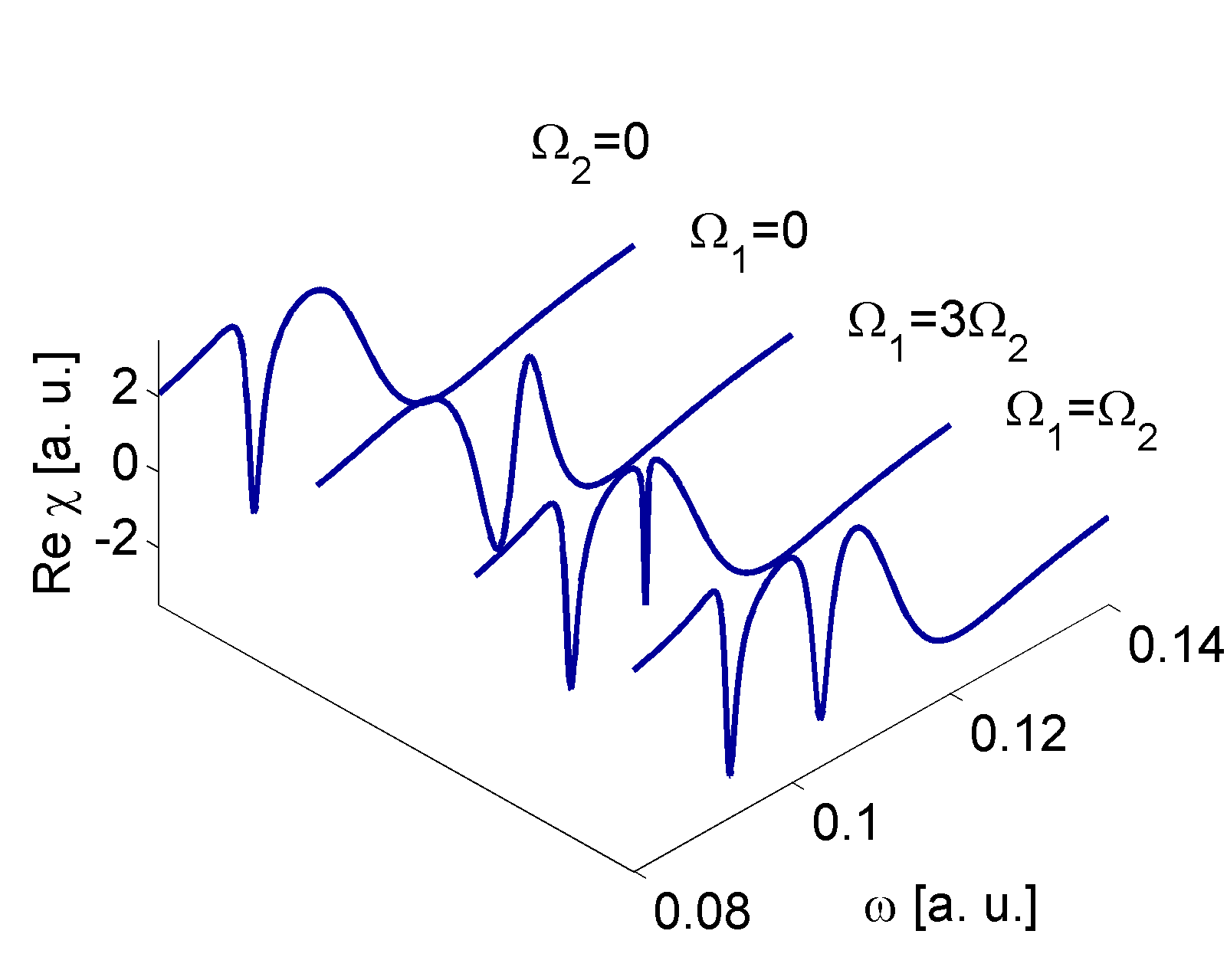} \caption{}
    \end{subfigure}
\caption{Real and imaginary part of $\chi(\omega)$ for the parameters $L=100$, $R_1=R_2=0.01$, $R_3=2$, $C_2=C_3=1$, $C_1=1.3$, with both coupling strengths equal ($C_{13}=C_{23}=10$), different ($C_{23}=3C_{13}$) or one of them disabled ($C_{13} \rightarrow \infty$ or $C_{23} \rightarrow \infty$)} \label{rys_dysp3d}
\end{figure}

\begin{figure}
\centering
\begin{subfigure}[b]{0.45\linewidth}
\includegraphics[width=1\linewidth]{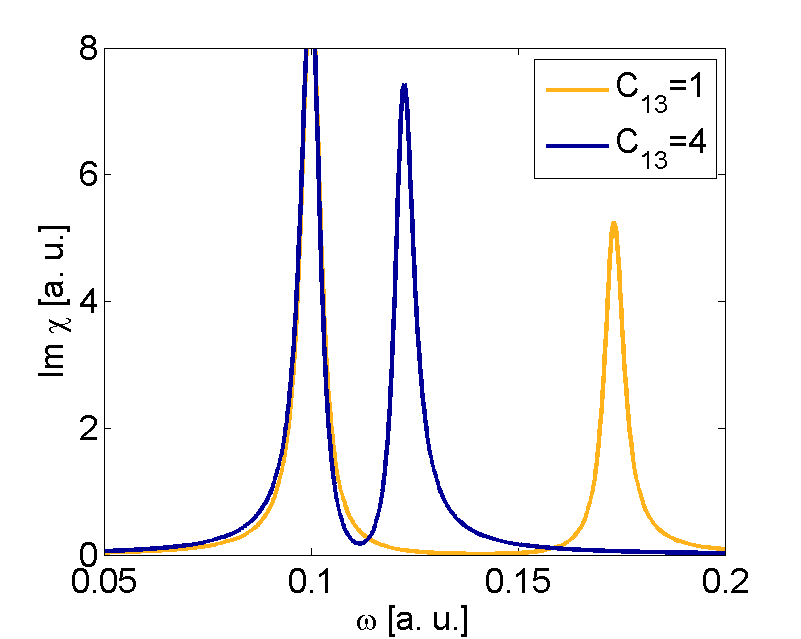}
\caption{} \label{rys_dysp1}
\end{subfigure}
\begin{subfigure}[b]{0.45\linewidth}
\includegraphics[width=1\linewidth]{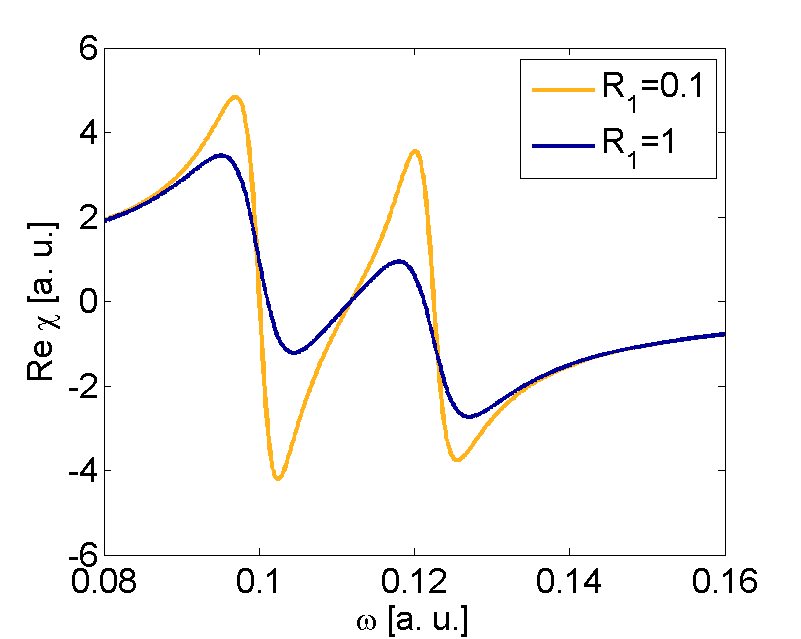}
\caption{}\label{rys_dysp2}
\end{subfigure}
\caption{a) Imaginary part of $\chi(\omega)$ for the parameters from Fig. \ref{rys_dysp}, $C_{23} \rightarrow \infty$ and two values of the coupling capacitance $C_{13}$. b) Real part of $\chi(\omega)$ for two values of $R_1$.}
\end{figure}

The most interesting property of a EIT system is the possibility of an active control over the transparency windows. On the Fig. \ref{rys_dysp3d}, one can see the real and imaginary part of the  susceptibility, featuring two transparency windows. When one of the control fields is disabled, a generic $\Lambda$ system with single window is obtained while for nondegenerated tripod system it is possible to get two transparency windows, the widths of which depend on the control fields strenghts.  From the Eq. \ref{atom_param}, one can see that in the electrical circuit model of tripod medium, the Rabi frequencies of the control fields $\Omega_{c1}$ and $\Omega_{c2}$ are functions which depend on capacitances and inductance. In order to examine the influence of coupling capacitance $C_{13}$ (or $C_{23}$) on the width of the transparency window, for a sake of simplicity we will consider the case $\Omega_{c2}=0$ which corresponds to $C_{23} \rightarrow \infty$, and a finite value of $C_{13}$. As the capacitance $C_{13}$ increases, the coupling $\Omega_{c1}$ becomes weaker, eventually vanishing in the limit of $C_{13} \rightarrow \infty$ meanwhile the transparency window becomes narrower and its position is  shifted to the lower frequency.  This is shown on the Fig. \ref{rys_dysp1}. It is justified by the fact that according to the Eq. \ref{param_elec}, both $\omega_1$ and $\omega_3$ are affected by the value $C_{13}$ which is responsible for coupling between two elements of the electric systems (see Fig. 1c). However, in the case of weak coupling and narrow window, $C_{13}>>C_1$ and $C_{13}>>C_2$, so that the frequency shift of the window is negligible. It is worth mentioning that the effect of down-shifting the position of transparency window due to increasing coupling capacitance was recently observed by Feng et al \cite{Feng}, but presented above considerations allow one to explain their observations.

In the electrical circuit model, the vanishing of the control fields $\Omega_{c1}$ and $\Omega_{c2}$ can be realized by removing or bypassing the capacitors $C_{13}$ and $C_{23}$. In the metamaterial model shown on the Fig. \ref{uklad}, the decoupling could be realized and controlled through the geometry by a significant increase of the distance $d_1$ and $d_2$.

 The dispersion relation is strongly affected by
the damping constants $\gamma_{bc}$ and $\gamma_{bd}$ and in the electric circuit damping is realized  by resistor elements (see Eq. \ref{param_elec}). In the realistic experiments multiple control schemes have been developed to overcome the losses. In order to minimize the Ohmic losses, the optical control by photoconductive  silicone islands  \cite{Gu2012} or temperature controlled superconductors \cite{Kurter} have been applied. As shown on the Fig. \ref{rys_dysp2}, the increase of absorption caused by change of the resistance $R_1$ significantly affects the slope of the dispersion curve inside the window, changing the group velocity which is given by
\begin{equation}
V_g = \frac{c}{1 + \frac{\omega}{2}\frac{\partial \chi'}{\partial \omega}},
\end{equation}
where dispersion $\chi'$ is a real part of complex  susceptibility. Our result confirmed the observation done by Kurter \emph{at al} \cite{Kurter} of a large enhancement in group delay, which enables a significant slowdown of the signal
propagating through the metal-superconductor hybrid metamaterial and demonstrates that pulse propagation  could be controlled on demand.

Another degree of control is realized by modification of the frequencies $\omega_1$, $\omega_2$, $\omega_3$. In particular, the temperature dependence of the carrier concentration in semiconductors can be used to alter the plasma frequency and shift the transparency window \cite{Bai2010}. In the presented circuit model, such a control corresponds to the change of the capacitances $C_1$, $C_2$, $C_3$.

\section{Numerical simulation of EIT metamaterial}
The Finite Difference Time Domain method has been used to simulate the pulse propagation through EIT metamaterial. A one-dimensional system has been assumed. The pulse consists of Gaussian envelope on a plane waves and travels along $\hat x$ axis. At every point of space, single electric field vector component $E_y$ and magnetic field component $H_z$ is defined. The usual field update formulas derived from Maxwell's equations \cite{Taflove} are complemented by the material response calculated with the Auxillary Differential Equations (ADE) method. The basis of the calculation are the time domain formulas presented in Eq.~\ref{diff_eq}. Assuming a unit scaling where $\epsilon_0=\mu_0=c=1$, one can write the Maxwell's equations in the form
\begin{eqnarray}
-\frac{\partial}{\partial x} E &=& \sigma_m H +\frac{\partial}{\partial t}\left(H + M\right),  \nonumber\\
-\frac{\partial}{\partial x} H &=& \sigma_e E + \frac{\partial}{\partial t}\left(E + P\right).
\end{eqnarray}
where $P$ and $M$ are the polarization and magnetization vector components and $\sigma$ are the conductivities. By introducing the notation
\begin{equation}
E(x\Delta x, n\Delta t) = E_x^n
\end{equation}
where $\Delta x$ and $\Delta t$ are finite spatial and time steps, one obtains the update equations
\begin{eqnarray}
H_{x+\frac{1}{2}}^{n+\frac{1}{2}} &=& \frac{2 - \sigma_m \Delta t}{2 + \sigma_m \Delta t} H_{x+\frac{1}{2}}^{n-\frac{1}{2}} - \frac{2\Delta t}{2 + \sigma_m \Delta t}\left[\frac{E_{x+1}^n - E_x^n}{\Delta x} + \frac{M_{x+\frac{1}{2}}^{n+\frac{1}{2}}-M_{x+\frac{1}{2}}^{n-\frac{1}{2}}}{\Delta t}\right], \nonumber\\
E_x^{n+1} &=& \frac{2 - \sigma_e \Delta t}{2 + \sigma_e \Delta t} E_x^{n} - \frac{2\Delta t}{2 + \sigma_e \Delta t}\left[\frac{H_{x+\frac{1}{2}}^{n+\frac{1}{2}} - H_{x-\frac{1}{2}}^{n+\frac{1}{2}}}{\Delta x} + \frac{P_{x}^{n+1}-P_{x}^{n}}{\Delta t} \right].
\end{eqnarray}
The medium is assumed to be magnetically inactive i. e. $M=0$. The oscillating charges described by Eq. \ref{diff_eq} give rise to three polarization values $P_1$, $P_2$, $P_3$. Only the third one is connected with the external field, so that $P=P_3=q_3d$, where $d$ is a constant length of the metamaterial structure. As with the electric and magnetic field, the polarizations are calculated using the first order differences
\begin{eqnarray}
P_i^n &=& \frac{4 - 2\omega_i^2\Delta t^2}{2 + \gamma_i \Delta t} P_i^{n-1} + \frac{-2 + \gamma_i\Delta t}{2 + \gamma_i \Delta t} P_i^{n-2} - \frac{2\Omega_i^2\Delta t^2}{2 + \gamma_i \Delta t}P_3^n, \qquad i=1,2\nonumber\\
P_3^n &=& \frac{4 - 2\omega_3^2\Delta t^2}{2 + \gamma_3 \Delta t} P_3^{n-1} + \frac{-2 + \gamma_3\Delta t}{2 + \gamma_3 \Delta t} P_3^{n-2} + \frac{2d^2\Delta t^2}{L(2 + \gamma_3 \Delta t)}E^n - \frac{2\Omega_1^2\Delta t^2}{2 + \gamma_1 \Delta t}P_1^n - \frac{2\Omega_2^2\Delta t^2}{2 + \gamma_2 \Delta t}P_2^n.
\end{eqnarray}
To ensure a satisfactory stability of the dispersion calculation scheme, for a spatial step $\Delta x=1$,
the time step was set to $\Delta t=0.5$. The frequency of the propagating pulse is such that its period $T \approx 50 \Delta t$, making numerical dispersion negligible \cite{Taflove}. The particular units of time and space are left as a free parameter, and they are connected by relation $c=\Delta t/\Delta x$.

It has been experimentally verified that EIT can be implemented  at optical frequencies at infrared region
 in  planar metamaterial gold strips, which were shifted on a distance of tens nm [16]. Moreover, it was should be kept in mind that trapping and retriving of electromagnetical waves of GHz regime has been performed in the metamaterial constructed of electric circuits structures with inductance of $L=180$nH and capacitances of order of pF \cite{Nakanishi}. Since all the medium parameters in Eq. \ref{param_elec} have units of frequency which is affected by scaling of $\Delta t$, our medium can be described by arbitrary, convenient numerical values of $R,L,C$. Therefore, the presented results are general and no reference to single, specific system is made.

\section{Simulation results}
\subsection{Verification of the medium model}

\begin{figure}[ht!]
\begin{subfigure}[b]{0.45\linewidth}
\includegraphics[width=\linewidth]{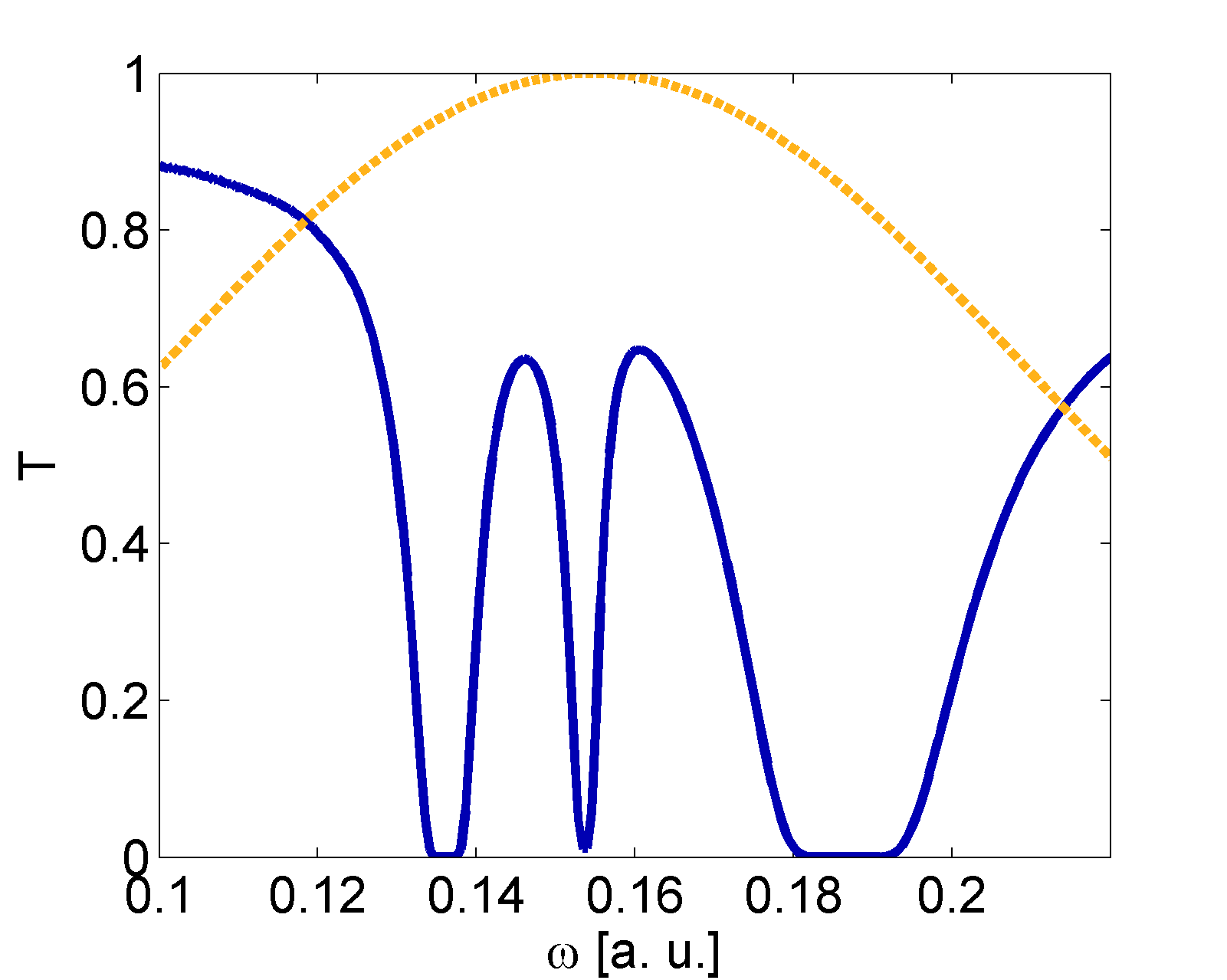}\caption{}
\end{subfigure}
\begin{subfigure}[b]{0.45\linewidth}
\includegraphics[width=\linewidth]{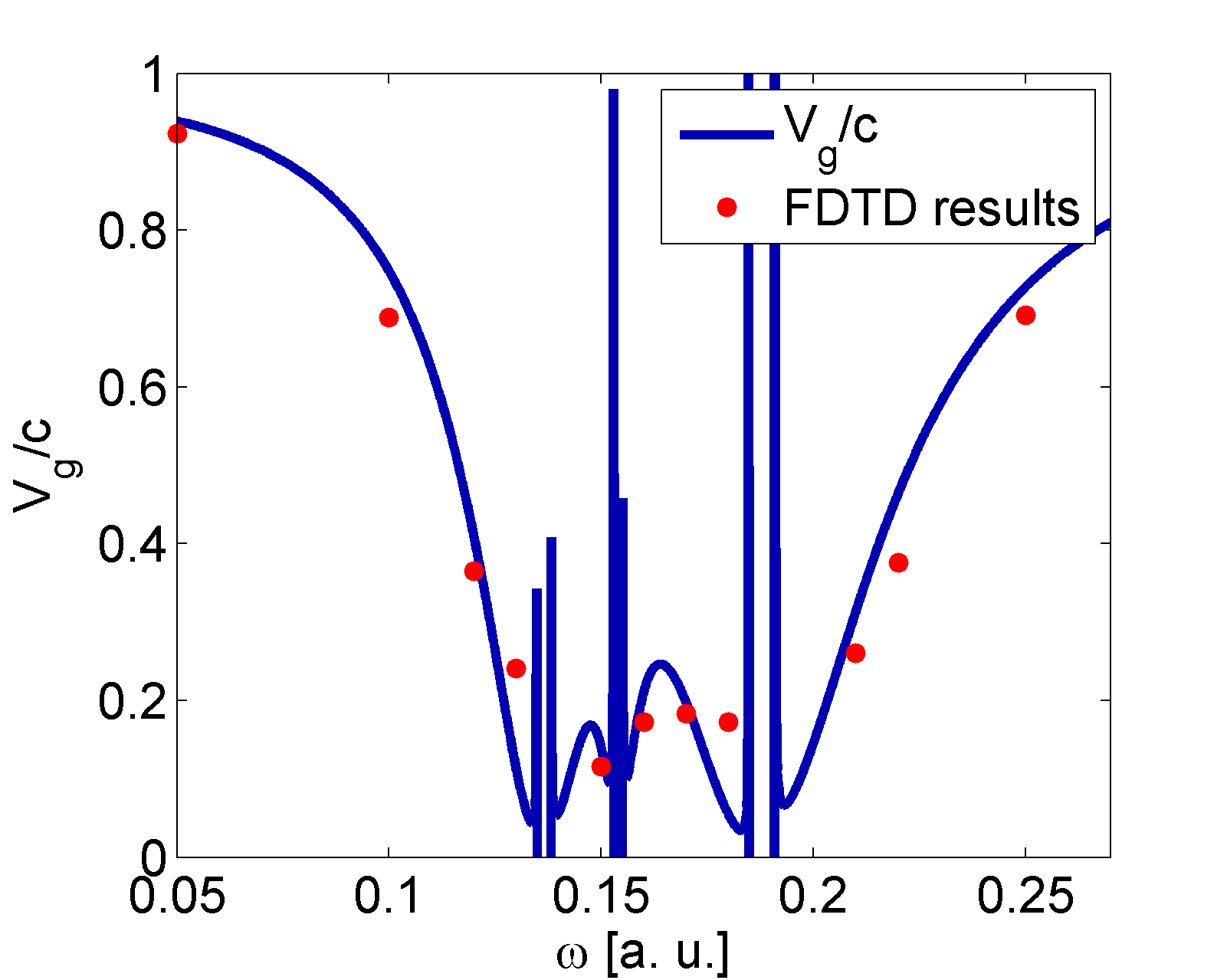}\caption{}
\end{subfigure}
\caption{a) Calculated transmission spectrum, with two easily visible transparency windows, incident pulse  is marked by a dashed line. b) The dependence of the group velocity on frequency inside the window. For $\omega = 0.145$ and $\omega=0.16$, $V_g \approx 0.2c$. }\label{Fig_verif}
\end{figure}

To verify the model, the propagation of a short, wide spectrum pulse through the metamaterial structure was simulated. The transmission coefficient based on the amplitude of the pulse entering and exiting the structure has been calculated. The results are shown on the Fig. \ref{Fig_verif} a. The pulse had a Gaussian envelope, with central frequency $\omega_0=0.17$, and its spectrum covered the whole region of interest in the dispersion curve. On the transmission spectrum, one can see two transparency windows with the transparency coefficient $T\approx0.6$ separated by regions with negligible transmission. Therefore, the simulation accurately  models the absorption spectrum of a tripod system. To test the dispersion of the medium, the dependence of the group velocity on the frequency has been investigated. Several pulses of various central frequencies have been generated, and their propagation time through the structure was obtained. The results are shown on the Fig. \ref{Fig_verif} b and one can see that the FDTD results are in a good agreement with theoretical value. Due to the finite spectral width of the pulse, small discrepancy appears in the regions where the group velocity changes rapidly with frequency.

\subsection{Simulation of signal storage}

Dynamic modulation of the EIT properties enables to slow, stop or even store the electromagnetical signal in the medium and after performing controlled  manipulations to retrieve it.
As it was described in previous sections, tripod medium offers richer possibilities to store and retrieve the the signal than the generic lambda system.
In order to illustrate it we perform the dynamic switching between EIT on and off regime, which enables one to store the signal in the EIT metamaterial and subsequently release it in two parts. The medium control is provided by modification of capacitances $C_{13}$ and $C_{23}$ which correspond  with   changing the Rabi frequencies $\Omega_{c1}$ and $\Omega_{c2}$ respectively. The parameters were set in such a way that in the initial state, when both control fields are active there are no detunings so that $\omega_1=\omega_2=\omega_3=\omega_0$, where $\omega_0$ is the central frequency of the pulse.

\begin{figure}[ht!]
    \centering
   \begin{subfigure}[b]{0.48\linewidth}
  \includegraphics[width=1\linewidth]{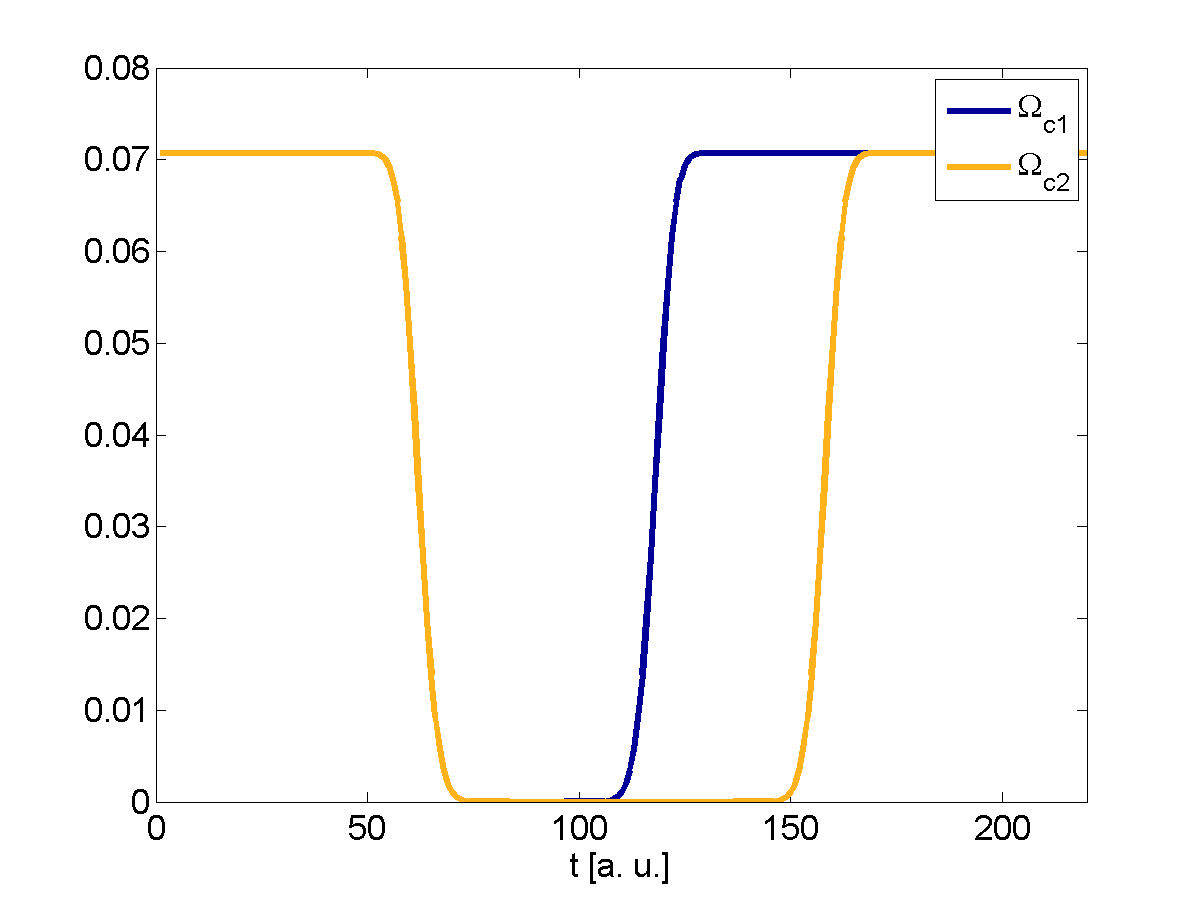} \caption{}
    \end{subfigure}
    \begin{subfigure}[b]{0.48\linewidth}
    \includegraphics[width=1\linewidth]{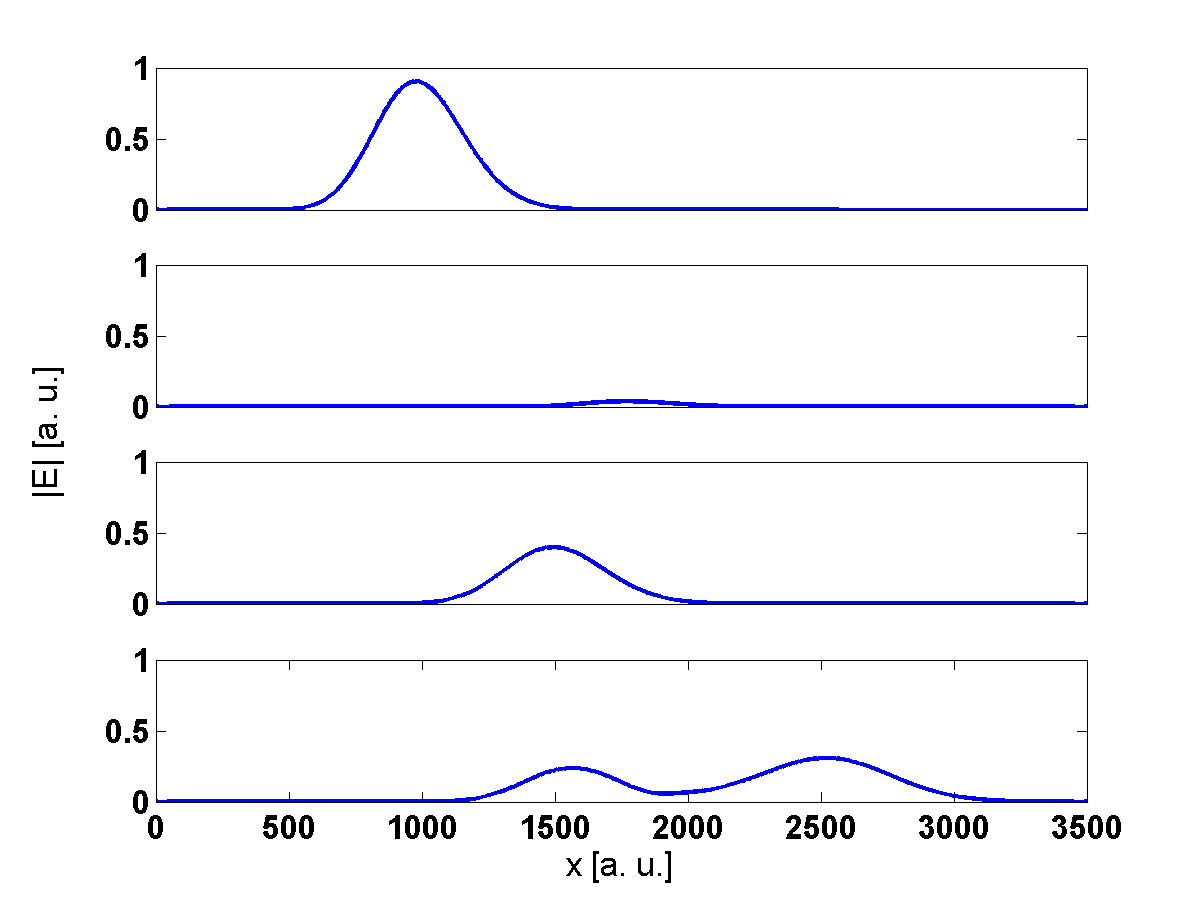} \caption{}
    \end{subfigure}
    \begin{subfigure}[b]{0.48\linewidth}
    \includegraphics[width=1\linewidth]{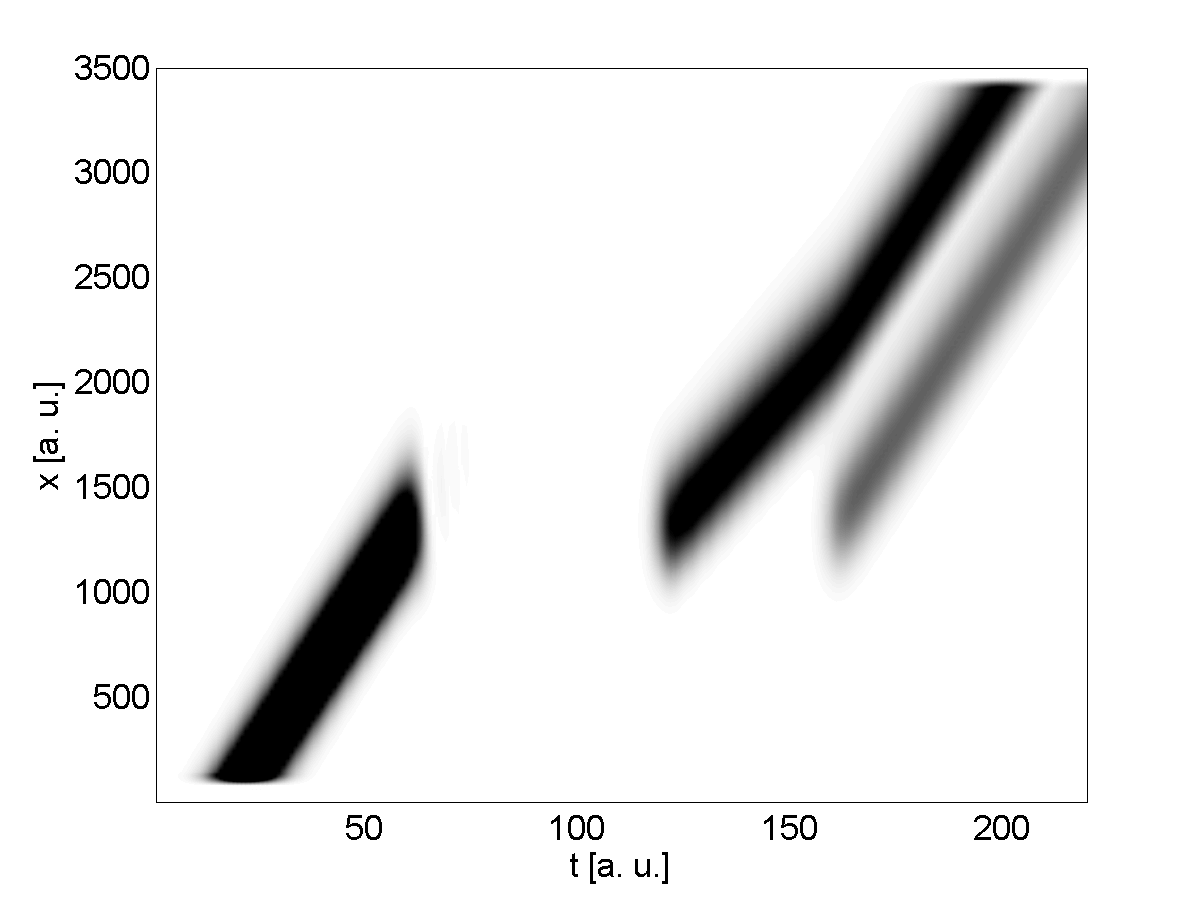} \caption{}
    \end{subfigure}
    \begin{subfigure}[b]{0.48\linewidth}
    \includegraphics[width=1\linewidth]{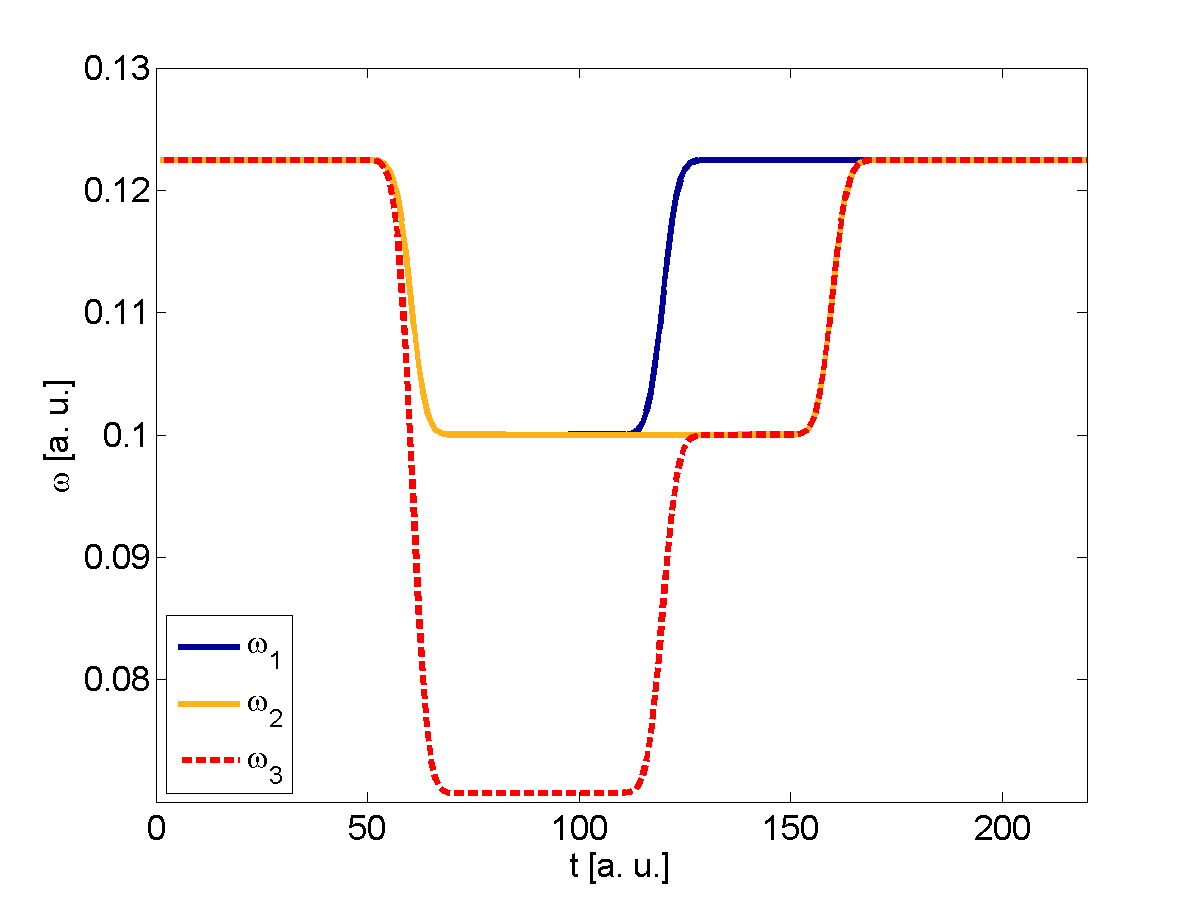} \caption{}
    \end{subfigure}
    \begin{subfigure}[b]{0.48\linewidth}
    \includegraphics[width=1\linewidth]{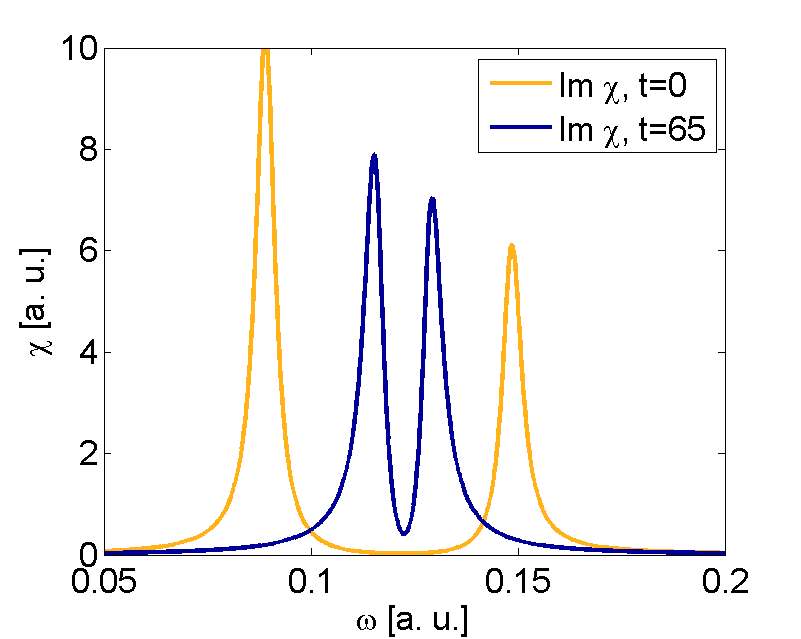} \caption{}
    \end{subfigure}
    \begin{subfigure}[b]{0.48\linewidth}
    \includegraphics[width=1\linewidth]{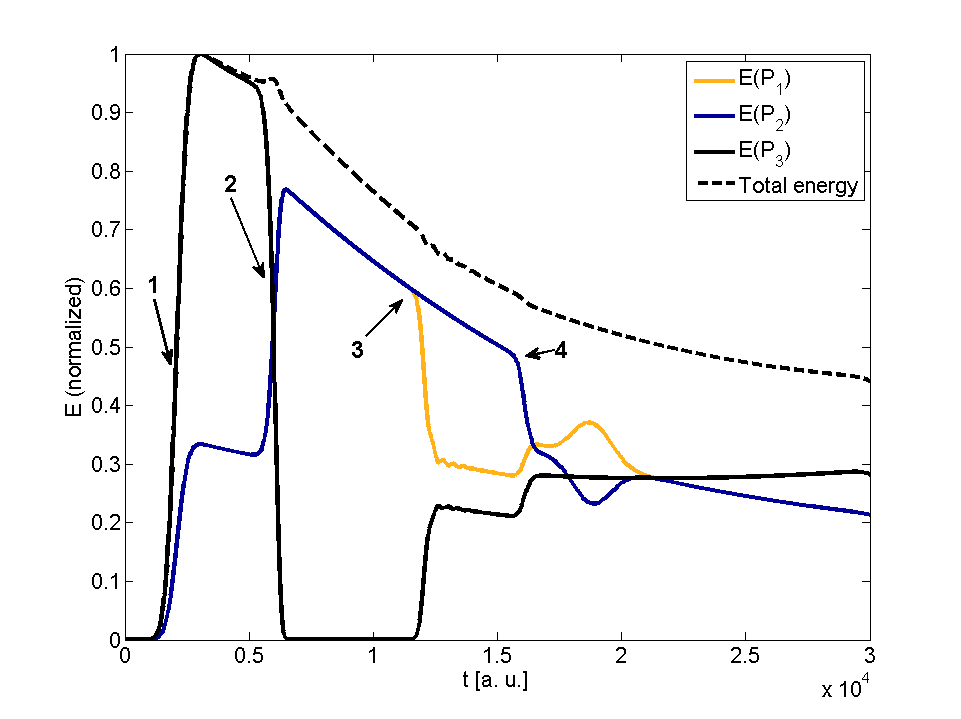} \caption{}
    \end{subfigure}
    \caption{Simulation results for $L=100$, $C_1=C_2=1$, $C_3=2$, $R_1=R_2=0.01$, $R_3=1$. a) Control field strengths as a function of time; b) Field snapshots illustrating the storage and release of the signal; c) Time and space dependence of the field strength, showing the storage process; d) Time dependence of the frequencies. e) Dispersion relation of the medium when constant values of $\omega_1$, $\omega_2$, $\omega_3$ are preserved. f) Normalized energy contained in polarizations $P_1$, $P_2$,  $P_3$ and external field.}\label{Res1}
\end{figure}

The control field strengths are shown on the Fig. \ref{Res1} a. They are equal at the storage
stage but the field 1 precedes the field 2 by some time at the
release stage. At $t=60$, both control fields decrease, vanishing completely at t=70. In terms of the circuit model, this corresponds to the situation where coupling capacitances increase significantly, transforming the system into three separate circuits. In the metamaterial model, such an effect is obtained by sufficient increase of the separation $d_1$ and $d_2$.
In the first snapshot on the Fig. \ref{Res1} b one can see the entering pulse at the left hand
side which is then stored inside the medium. After the storage stage, when the pulse energy is confined inside the metamaterial as electric charge oscillations denoted by currents $i_1$ and $i_2$ the signal is released in two portions, i.e. at $t=110$, first control signal is turned on again, releasing the first part of the pulse,  similarly as in the simple $\Lambda$ system. The stored energy starts to be radiated when the transparency window begins to open. Finally, the second part is released at $t=150$ by increasing $\Omega_{c2}$.  The process is illustrated on the Fig. \ref{Res1} c, where time and space dependence of the signal strength is presented. However, the latter releasing occurs in the presence of both control fields, which gives rise to a difference of the two parts of the signal, as concerns their heights and initial velocities, which is clearly seen in Fig. \ref{Res1} c. The first part has been released
with a zero initial velocity  while the second part has a nonzero velocity from the very beginning because two control fields are on. To obtain a full symmetry one should switch the first control field off before switching the second one on. Then one would obtain two identical released pulses shifted in time, as in two independent simple $\Lambda$ systems.

One can see that a significant fraction of the pulse is restored, but some leaked signal continues to propagate through the medium when the control fields are off (see Fig. \ref{Res1} b).
 The point is that due to dynamical changes of $C_{13}$ and $C_{23}$  at the time when both control fields $\Omega_{c1}$ and $\Omega_{c2}$ are switched off, the significant shifts of frequencies $\omega_1$, $\omega_2$ and $\omega_3$ appeared (see Eq. 4) and, as it was pointed out in Sec. II B, the shape of the window is distorted, window becomes narrow and is shifted to the lower frequencies.
  As a consequence, at some point during the storage process, when the coupling field strength is weak, the central frequency of the pulse is outside of the narrow window and, as a result, some part of the signal is not absorbed.

 The shift of frequencies $\omega_1$, $\omega_2$ and $\omega_3$ can be minimized when the coupling capacitances $C_{13}$ and $C_{23}$ are much bigger than $C_1$, $C_2$ and $C_3$. In principle, one could also dynamically modify these values so that despite the changing values of $C_{13}$ and $C_{23}$, the frequencies $\omega_1$, $\omega_2$ and $\omega_3$ may remain constant. For example, basing on Eq. \ref{param_elec} to obtain $\omega_1=const$, one has to change the value of $C_1$ according to the formula
\begin{equation}
C_1 = \left(L\omega_1^2 - \frac{1}{C_{13}}\right)^{-1}.
\end{equation}
Analogous procedure can be repeated for $C_2$ and $C_3$. Simulation results for such a case are shown on the Fig. \ref{Res1} e. One can see that the symmetry of the window is restored. Therefore, a full analogue of atomic tripod system realized in a metal strip metamaterial shown in Fig. 1b requires control over multiple parameters. For instance, one could combine the mechanical change of distances $d_1$ and $d_2$, affecting the coupling $\Omega_1$ and $\Omega_2$, and the temperature dependence of the frequencies $\omega_1$, $\omega_2$ and $\omega_3$ \cite{Bai2010}.
Finally, the Fig. \ref{Res1} f depicts the normalized measure of the energy stored in the system. In analogy to the mechanical oscillator model, we can calculate the energy
\begin{eqnarray} \label{Energie}
E_1 &=& \frac{1}{2}L\omega_1^2q_1^2 + \frac{1}{2}L \dot{q_1}^2, \nonumber\\
E_2 &=& \frac{1}{2}L\omega_2^2q_2^2 + \frac{1}{2}L \dot{q_2}^2, \nonumber\\
E_3 &=& \frac{1}{2}L\omega_3^2q_3^2 + \frac{1}{2}L \dot{q_3}^2 + q_3V_{ext} + E^2,
\end{eqnarray}
stored in the medium polarizations, where the first term is the potential energy and the second term is the kinetic energy. In terms of RLC circuit, these two factors correspond to the energy stored in capacitor and induction coil, respectively. Moreover, the part $E_3$ contains also the energy of field-polarization interaction $q_3V_{ext}$ and the vacuum field energy $\epsilon_0 E^2$, where $\epsilon_0=1$.

One can see that the energy reaches the maximum value when the pulse enters the system (Fig. \ref{Res1} f, 1). Then, when the window is closed, it is stored in the form of polarizations $P_1$ and $P_2$ (2). The points (3) and (4) mark the moments when one of the control fields is turned on and one part of the pulse is released. Interestingly, when the second control field is switched of, the energy oscillates between $P_1$ and $P_2$ for some time. However, the total energy is conserved. As expected, the total energy decreases exponentially. Some transient effects are visible at the points when the window is opening or closing, indicating that the measure given by Eq. \ref{Energie} is sufficient only for a steady state.

\subsection{Train of pulses}
\begin{figure}[ht!]
\begin{subfigure}[b]{0.48\linewidth}
\includegraphics[width=\linewidth]{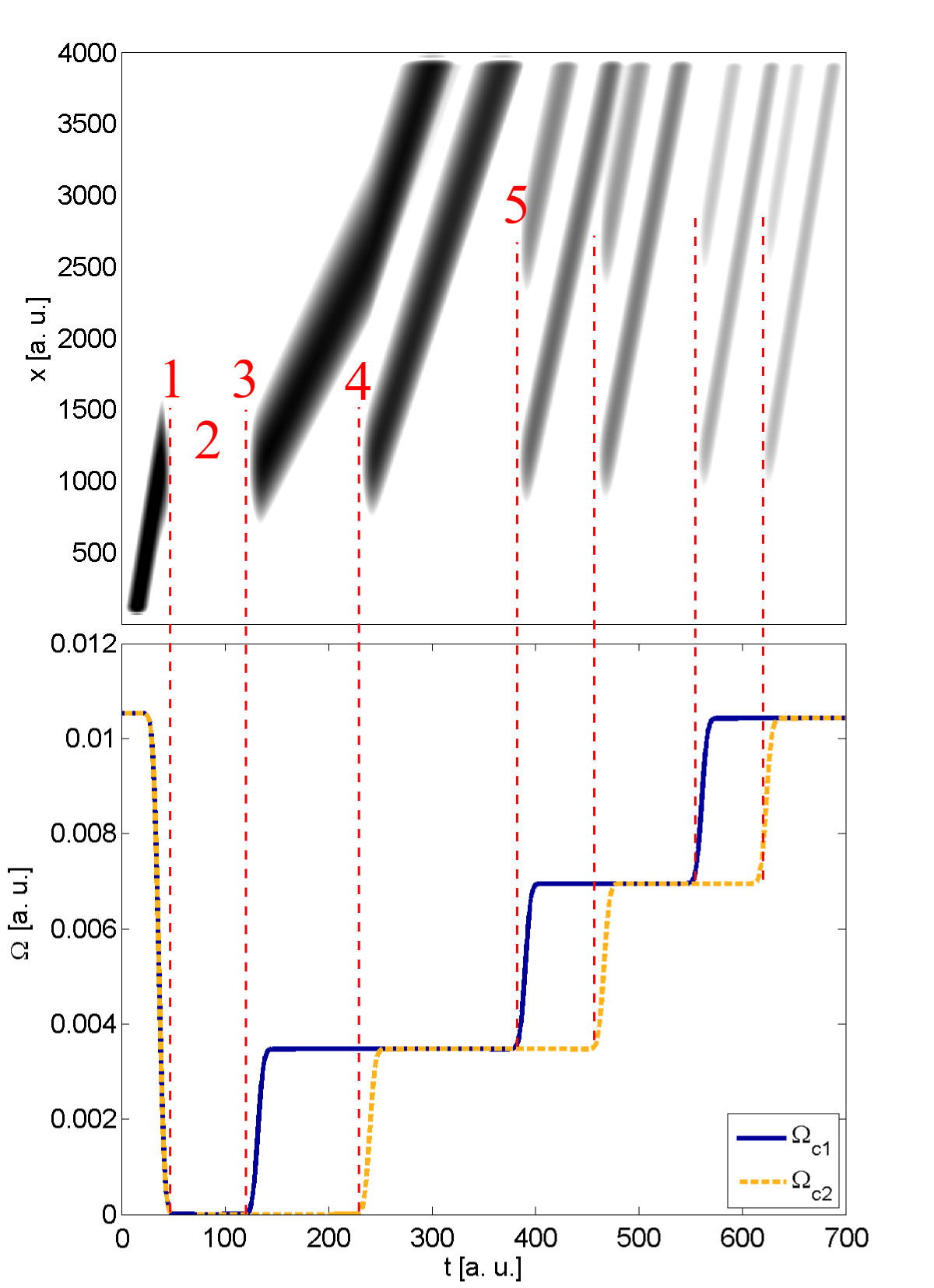}
\end{subfigure}
\begin{subfigure}[b]{0.48\linewidth}
\includegraphics[width=\linewidth]{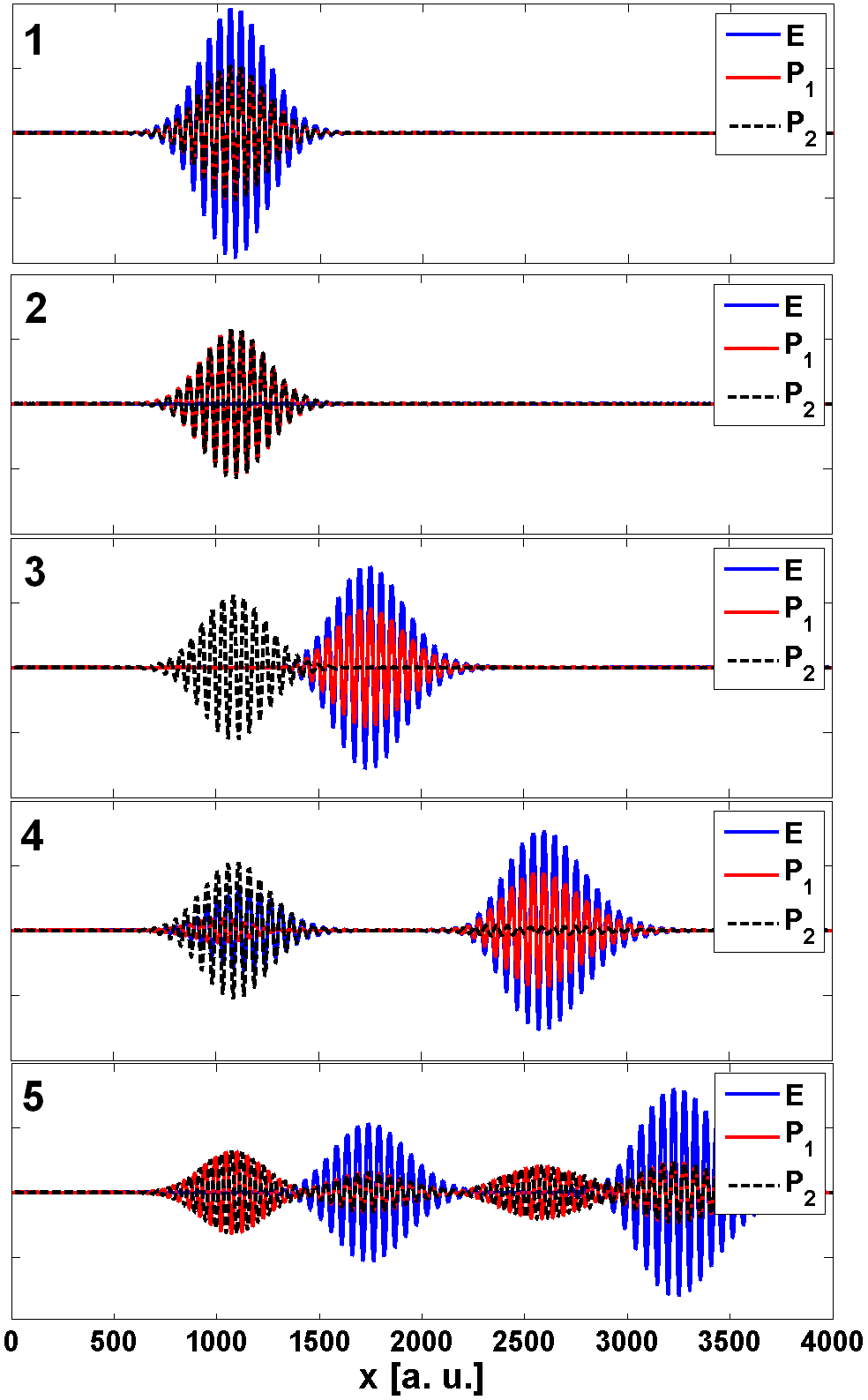}
\end{subfigure}
\caption{Simulation results for storge and release of the signal in many parts, forming a train of pulses. a) Time and space dependence of the pulse amplitude along with the control field strengths, illustrating the storage of the pulse and its release in multiple stages. b) Field snapshots at the specific times, showing both the propagating and stored part of the signal.}\label{Fig_chain}
\end{figure}

  It is possible to release the stored signal in a form of multiple subsequent pulses by increasing the coupling strengths in multiple steps at the releasing. The simulation results for such a case is presented on the Fig. \ref{Fig_chain}. To better understand the dynamics of the storage process, field snapshots have been taken at the characteristic moments. On the first panel of the Fig. \ref{Fig_chain} b, one can see the initial, propagating pulse which consists of the electric field $E$ and the two polarizations $P_1$, $P_2$ coupled by $\Omega_{c1}$ and $\Omega_{c2}$. When the control fields are disabled (Fig. \ref{Fig_chain}a, 1), the pulse is stored inside the medium in the form of localized oscillations of $P_1$ and $P_2$. When the first control field is turned on, the polarization $P_1$ becomes coupled to the external field $E_{ext}$. As a consequence, a propagating pulse is formed (third panel). Then, when the amplitude of the second control field is increased, another pulse is generated by using the energy stored in polarization $P_2$. At the same time, the first pulse also becomes coupled to $P_2$, forming a new, localized perturbation of $P_1$ and $P_2$. As a result, further changes of one of the control fields generate two pulses (fifth panel). This is easily visible on the Fig. \ref{Fig_chain} a, where the last four changes of the control fields generate pair of pulses each. The positions of these pulses correspond to the points where the initial released pulse was located when the first and second control fields were switched on (Fig. \ref{Fig_chain}a 3, 4).

\section{Conclusions}
We have considered the
details of EIT and the dynamics of pulse propagation in a classical analogs of tripod system such as electric circuit or properly arranged metal strips
and  some quantitative predictions concerning the characteristics
of a pulse/pulses stored in such media have been presented and discussed both analytical and numerical in terms of the energy and polarization of the system. Both considered classical models of the tripod structure allow one for steering the propagation through different combinations of coupling capacitors or geometrical changes. Our theoretical and numerical results  confirm and explain recently observed effect of the dependence of transparency window position on coupling capacitances. Due to rich dynamics and controllability, the tripod medium allows for a flexible and effective processing of the stored signal and its release on demand in one or more parts with prefect control of their intensity.  The performed FDTD simulations confirm the close analogy between atomic tripod system and its classical, metamaterial counterpart and provide an insight into the dynamics of the signal processing. Moreover,  slow-light techniques realized in semiclassical media and solid state metamaterials   hold great promise for applications in telecom and quantum information processing.

\end{document}